\documentclass[12pt]{article}
\usepackage[pdftex,colorlinks=true,linkcolor=blue,citecolor=blue,urlcolor=blue]{hyperref}
\usepackage[usenames,dvipsnames]{xcolor}
\usepackage{amsmath,amsfonts,amssymb}
\usepackage{graphicx}
\usepackage{psfrag}
\usepackage{enumerate}
\usepackage{mathrsfs}
\usepackage{dsfont}
\usepackage{textcomp}
\newcommand\blfootnote[1]{%
  \begingroup
  \renewcommand\thefootnote{}\footnote{\hspace{-6mm}#1}%
  \addtocounter{footnote}{-1}%
  \endgroup
}

\def\({\left(}
\def\){\right)}
\def\[{\left[}
\def\]{\right]}

\numberwithin{equation}{section}

\begin{document}
%\noLabels % uncomment for final production
%\nobbibitem % uncomment for final production
\clearpage\thispagestyle{empty}

\begin{center}

%\begin{flushright} \vspace{-3cm}
%{\small UPR-1222-T}  \\
%\end{flushright}
%\vspace{1cm}

{\large \bf
State-Dependent Divergences in the Entanglement Entropy}

\vspace{7mm}

Donald Marolf$^{a}$, and Aron C. Wall$^{b}$  \\

\blfootnote{\tt marolf@physics.ucsb.edu, aroncwall@gmail.com}

%\\

%\vspace{5mm}

\bigskip\centerline{$^a$\it Department of Physics, University of California,}
\smallskip\centerline{\it Santa Barbara, CA 93106, USA}
%\bigskip\medskip
\bigskip\centerline{$^b$\it School of Natural Sciences, Institute for Advanced Study}
\smallskip\centerline{\it Princeton, NJ, USA}
%\bigskip\medskip
%\vfil

\end{center}

\vspace{5mm}

\begin{abstract}
We show the entanglement entropy in certain quantum field theories to contain state-dependent divergences.  Both perturbative and holographic examples are exhibited.  However, quantities such as the relative entropy and the generalized entropy of black holes remain finite, due to cancellation of divergences. We classify all possible state-dependent divergences that can appear in both perturbatively renormalizeable and holographic covariant $d\le 6$ quantum field theories.
\end{abstract}

\setcounter{footnote}{0}
\newpage
\clearpage
\setcounter{page}{1}

\tableofcontents

%%%%%%%%%%%%%%%%%%%%%%%%%%%%%%%%%%%%%%%%
%
\section{Introduction}
%
%%%%%%%%%%%%%%%%%%%%%%%%%%%%%%%%%%%%%%%%
\label{sec:intro}

There has been much recent interest in the entanglement entropy of quantum field theories (QFTs).  Given a QFT, a state in the theory, and a choice of region $R$, the entanglement entropy is formally defined as
\begin{equation}\label{vN}
S_R = -\mathrm{tr}(\rho_R \ln \rho_R),
\end{equation}
where $\rho_R$ is the restriction of the state to the region $R$.  This quantity is sensitive to all degrees of freedom in $R$ (in fact, it is invariant under unitary transformations $\rho_R \to U\,\rho_R\,U^{-1}$) and obeys a set of interesting inequalities \cite{wehrl78}. It is related to c-theorems in various dimensions \cite{Casini:2004bw,Casini:2006es,Casini:2012ei,Grover:2012sp,Solodukhin:2013yha}, and in certain cases it provides information about topological phases which cannot be obtained from local order parameters \cite{Kitaev:2005dm,LW06,Grover:2013ifa}.  It is also of interest in holographic theories, where there is a simple geometric description in terms of the area of an extremal surface in the dual bulk theory \cite{Ryu:2006bv,Ryu:2006ef,Hubeny:2007xt}.

However, in any unitary theory with local degrees of freedom, $S_R$ is divergent.  The leading order divergence is proportional to the area $A$, but there may also be subleading divergences.  This makes Eq. (\ref{vN}) a purely formal expression until a UV regulator is specified, e.g. a lattice\footnote{ Cf. \cite{Buividovich:2008gq,Donnelly:2011hn,Casini:2013rba} for subtleties involving gauge theories.}, brick wall \cite{hooft1985quantum}, Pauli-Villars \cite{Demers:1995dq}, heat kernel (reviewed in \cite{Solodukhin:2011gn}), mutual information regulator \cite{Casini:2004bw,Iqbal:2015vka}, etc.  In general, the value of $S_R$ will depend on the choice of regulator, making the interpretation of $S_R$ somewhat subtle.  Nevertheless, one simplifying feature is that the divergences are always local integrals of quantities defined on the boundary entangling surface $\partial R$.  For example, in a $d$-dimensional scale-invariant theory with a regulator at an energy scale $\Lambda$, one often finds power laws and/or logarithmic divergences:\footnote{In non-scale-invariant theories such as a free massive scalar, certain simple regulators produce more exotic divergences such as $\log \log \Lambda$ \cite{DWforthcoming}.}
\begin{equation}
S_R = \sum_{n < d} \int_{\partial R} X^{(n)}\,dA\,\Lambda^{D-2-n} + \int_{\partial R} X^{(d)}\,dA\,\ln(\Lambda) + \text{finite},
\end{equation}
where $X^{(n)}$ are local integrands of dimension $n \le d$.  In cases where each $X$ is constructed out of geometrical structures such as intrinsic and extrinsic curvatures, these dimensions are integers.\footnote{If we further assume that $X$ depends covariantly only on the metric tensor expanded near $\partial R$, then these integers must be even.  Subject to these assumptions, log divergences can only appear in even dimensions.}  But more generally $X$ can also depend on scale-dependent parameters in the theory such as masses, or even---as we shall argue below---on expectation values of quantum fields.  If these sources or fields have anomalous dimensions,  then $n$ will generally be non-integer.

There are several distinct strategies for dealing with this divergence, depending on the needs of your particular application:
\begin{enumerate}
\item \textbf{Specialize:} Accept that each different regulator defines a distinct quantity, so just pick one choice of regulator and stick to it.  (This is a somewhat narrow viewpoint because it makes it difficult to compare different calculations, but it may be fine within a particular project.)

\item \textbf{Universalize:} Identify ``universal'' aspects of $S_R$ which are the same for every good regulator (although they may still depend on the theory or state).  This includes the coefficients of log divergences, or the finite piece modulo a local counterterm \cite{Ryu:2006bv,Ryu:2006ef,Casini:2006ws,DWforthcoming}

Other sets of universal quantities related to entanglement are the mutual information $I_{A,B} = S_A + S_B - S_{AB}$ between two regions $A$ and $B$ separated by a finite spatial separation \cite{Casini:2006ws}, and the relative entropy of two states $S(\rho\,|\,\sigma) = \mathrm{tr}(\rho (\ln \rho - \ln \sigma))$ \cite{araki1976relative}.  These quantities are typically finite.

\item \textbf{Renormalize:} In a gravitational theory, black holes have finite entropy $S_{BH} = A/{4\hbar G} + \text{subleading\,corrections}$. Divergences in the QFT state outside the horizon can be handled by absorbing them into renormalizations of the gravitational parameters, e.g. the area law divergence corresponds to a shift of $1/G$, so that the total ``generalized entropy' $S_\text{gen} = S_\text{BH} + S_\text{ent}$ remains finite.  (See the Appendix of \cite{Bousso:2015mna} for a review and references.)  Although this approach was originally designed for black holes and other causal horizons, at least semiclassically one can assign an entropy to more general choices of entangling surface $\partial R$ as well \cite{C:2013uza,Myers:2013lva,Engelhardt:2014gca}.

\item \textbf{Subtract:} If the divergences do not depend on the state, then one can subtract off the entropy of some reference state $\sigma$, e.g. the vacuum state, as done in \cite{Holzhey:1994we,Marolf:2003sq,Casini:2008cr,Bousso:2014sda,Bousso:2014uxa}  (This is obviously not useful if you are only interested in the structure of vacuum entanglement entropy!)
\end{enumerate}

The purpose of this article is to identify situations in which the divergences depend on the state via the expectation value of some operator $\langle \mathcal{O} \rangle$.  In such examples, the vacuum subtraction approach to controlling divergences will fail, although the other three approaches remain valid.

One might have thought that any two reasonable states (having the same UV vacuum structure) should differ by only a finite entropy in any region, but this turns out not to be the case.  Physically reasonable states do indeed have finite relative entropy and/or generalized entropy, but in each of these cases there is an additional term in the expression (the modular Hamiltonian $\mathrm{tr}(\rho \ln \sigma)$ or a ``Wald-like'' surface term \cite{Wald:1993nt,Jacobson:1993vj,Iyer:1995kg,Faulkner:2013ana} respectively).   In the simplest examples, the divergences which cancel between the two terms are state independent so that $\Delta S_R = -\mathrm{tr}(\rho \ln \rho) + \mathrm{tr}(\sigma \ln \sigma)$ is well defined.  But in more complicated examples, $\Delta S_R$ has a divergence, which is nevertheless cancelled by the remaining part of the relative/generalized entropy, as discussed below.\footnote{Note that the finiteness of relative entropy occurs only for physically reasonable QFT states.  Even in a system as simple as a single harmonic oscillator  and taking $\sigma$ to be the thermal state, there exist normalizable states $\rho$ whose probability falls off sufficiently slowly with energy to make the relative entropy
$S(\rho | \sigma)$ diverge.  The relative entropy can also be infinite if there exist projections on which $\sigma$ has 0 probability, although when $\sigma$ is a vacuum state restricted to a compact region, the Reeh-Schlieder property tells us that no such projection operators exist.}

In order to keep the analysis under control, we will focus our attention on perturbatively renormalizeable or holographic theories with spacetime dimension $d\le 6$.  Thus we allow the theory to have marginal or relevant couplings, as long as these couplings are covariant.  In other words, if we start with the CFT that describes physics in the UV, then the only sources we may add are those given by the spacetime metric and constant scalars.  We will classify all state-dependent divergences that can appear in such theories.   In the CFT case, we argue that our divergences are generic by using bottom-up holographic examples defined by positing a dual asymptotically AdS bulk description.  While there are no state-dependent divergences in the simplest possible cases (free theory and source-free holographic CFT's), going beyond these assumptions allows us to find examples.

A subtlety, however, is that not all terms superficially allowed by dimensional analysis can appear in the entanglement entropy.  An important consistency principle comes from the ``replica trick''\footnote{reviewed in \cite{Calabrese:2009qy}, Cf. section A.1 of \cite{Bousso:2015mna} for additional references.}, which is a relationship between the entropy and the effective action.  Consider QFT states which are defined by some Euclidean path integral defined on a manifold $M$.  Given a choice of entanglement surface $\partial R$, we can define an $n$-sheeted replica manifold $M^{(n)}$ by introducing a conical singularity at $\partial R$ with total angle $2\pi n$, so that away from $\partial R$ the manifold is copied exactly $n$ times.  Let the partition function on this manifold be $Z^{(n)}$.  Assuming we can analytically continue to real valued $n$, the entropy is given by:
\begin{equation}\label{replica}
S_R = (1 - \partial_n) \ln Z^{(n)} |_{n = 1}.
\end{equation}
This tells us that local divergences in the entropy must be associated with the counterterms in the effective action of the QFT (at least if the UV regulator respects Eq. (\ref{replica})) \cite{Larsen:1995ax}.  We will make free use of such constraints in what follows.  In order for a term in the action to contribute, it must involve the Riemann curvature tensor, so that there is a nontrivial contribution coming from the tip of the conical singularity.  (Away from the conical singularity, local divergences drop out due to being linear in $n$.)  A further consequence is that the dependence of $S_R$ on the extrinsic curvature $K_{ab}$ of $\partial R$ is determined relative to other terms in the entropy functional \cite{FPS, Dong:2013qoa,Camps:2013zua}.

We begin with perturbations around free theories in section \ref{pert}.  Here our classification is performed simultaneously with the construction of examples exhibiting state-dependent divergences.  Section \ref{strong} then proceeds to classify possible state-dependent divergences at leading order in large $N$ for marginal or relevant deformations of $d \le 6$ holographic conformal theories; i.e., those with dual descriptions in terms of the classical gravitational dynamics of asymptotically AdS bulk spacetimes.  In this case we save the construction of examples for separate treatment in section \ref{holo}.  These examples are constructed in a bottom-up manner on the gravitational side of the duality.  In section \ref{finite} we explain how quantities such as the mutual information, generalized entropy, and relative entropy can remain finite even when the entanglement entropy has state-dependent divergences.  We close with some final discussion in section \ref{disc}.

%%%%%%%%%%%%%%%%%%%%%%%%%%%%%%%%%%%%%%%%
%
\section{Perturbatively Renormalizable Theories}
%
%%%%%%%%%%%%%%%%%%%%%%%%%%%%%%%%%%%%%%%%
\label{pert}

Implementing the replica trick perturbatively involves evaluating Feynman diagrams on the replica manifold $M^{(n)}$.  Assuming the spacetime metric to be smooth, obtaining a state-dependent divergence from a path integral requires a Feynman diagram with three properties:

\begin{enumerate}[a)]
\item it contains at least one loop (to make it divergent)
\item it has external legs ending on the entangling surface (to make it state-dependent)
\item it renormalizes a term in the action involving curvature (in order to contribute to the replica trick)
\end{enumerate}

In the case of free field theory, since there are no nontrivial vertices, a connected Feynman diagram cannot have both loops and loose ends.  So state-dependent divergences are forbidden in regular states of a free theory.\footnote{Cooperman and Luty \cite{Cooperman:2013iqr} claim to have found states with state-dependent divergences in free field theories.  However, these states were constructed by a path integral on a Euclidean manifold $M'$ which differed from (the Wick rotation of) the manifold $M$ on which the states were defined to live, and in particular where $M$ and $M'$ do not match smoothly.  States generated from this construction are in general not guaranteed to be regular states, for example they need not obey the Hadamard condition \cite{kay1991theorems,wald1994quantum}.}

\begin{figure}[ht]
\includegraphics[width = \textwidth]{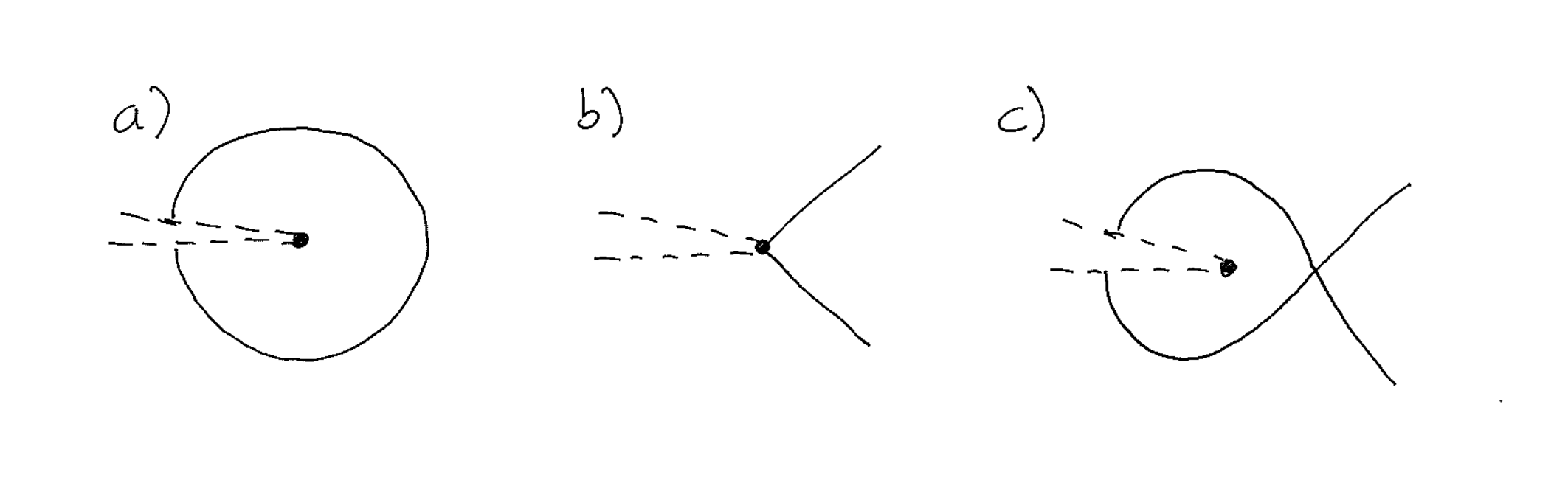}
\caption{\footnotesize Evaluating the entropy by the replica trick requires calculating the partition function $Z$ on a Euclidean spacetime manifold which is shaped like a cone of angle $\beta = 2\pi n$ near the entangling surface.  (In the case where there is rotational symmetry, one can easily analytically continue to noninteger $n$ values, and for ease of visualization we are showing the case where $n$ is slightly less than 1 so that a small angle is cut out of the plane normal to $R$; however similar diagrams exist for integer $n$.  Here the central dot is $\partial R$ and we are suppressing all but 2 dimensions.  Above are shown position-space Feynman diagrams which provide contributions to $\ln Z$.  a) A divergent diagram which contributes to the entanglement entropy $S_R$, which is state-independent because there are no external lines.  Its counterterm is a purely geometric functional of the boundary $\partial R$.  b) In a nonminimally-coupled theory with a $\phi^2 {\cal R}$ term, there is a contact interaction where the conical singularity couples directly to $\phi^2$.  This diagram provides a state-dependent finite contribution to the boundary Wald entropy term $S_{\partial R}$.  (There is an associated state-independent divergence in the Wald entropy if the two external lines are contracted into a loop.)  c) Here at last is a state-dependent divergence, which involves a single quartic interaction and two external lines.  This is a divergence in $S_R$, but the counterterm involves contracting the loop to a point to produce diagram (b), resulting in a nontrivial RG flow of the nonminimal coupling term.  Thus the renormalized sum $S_\text{gen} = S_R + S_{\partial R}$ is finite.}
\label{fig:conicalS}
\end{figure}

On the other hand, it is easy to generate such diagrams in interacting theories.  For example, in a $\phi^4$ theory with $d=4$, heat kernel methods give a logarithmically divergent counter-term in the action proportional to the integral of $\phi^2 {\cal R}$.  Such curvature couplings are well-known to contribute an entropy term proportional to the integral of $\phi^2$, here with logarithmically divergent coefficient.  A similar result may also be obtained by noting the mass-dependence of the logarithmic entropy divergence found in \cite{Rosenhaus:2014ula}, and that linearizing the $\phi^4$ term about states with non-zero expectation values of $\phi^2$ generally shifts the effective mass by an amount that depends on the choice of such a state\footnote{We thank Vladimir Rosenhaus for suggesting this point of view.}.  See Fig \ref{fig:conicalS} for an explanation of this state-dependent divergence in terms of Feynman diagrams.

\subsection{Classification of Divergences}
\label{pertclass}

We now classify all possible state-dependent divergences in perturbatively (superficially) renormalizable field theories in every dimension.  We emphasize that we treat these theories in a perturbative manner, so we allow unstable theories, and theories with Landau poles.  All divergences are either power laws or log divergences.\footnote{Thus we do not perform a one-loop resummation to calculate the beta function.  If we did, then theories like $d = 4$ Yang-Mills with a logarithmically running coupling constant would have a more complicated divergence structure.}  We allow covariant (constant scalar) sources only, though the analysis does not significantly change if one also allows coupling to a background gauge field.

As in the example above, state-dependent divergences correspond to renormalizations of nonlinear coupling terms such as $\xi \phi^2 {\cal R}$.  However, we do not consider the effects of bare nonminimal coupling terms added directly to the action, i.e. we do not calculate state-dependent divergences in the entropy that are themselves proportional to powers of $\xi$.  These terms produce a nontrivial Wald entropy term associated with the black hole entropy, arising from the conical singularity in $R$ at the entangling surface, e.g. in this case a term proportional to $-\xi \langle \phi^2 \rangle$.  Such terms in the Wald entropy may have divergences which renormalize other terms in the Wald entropy \cite{Donnelly:2012st}, but are not normally considered to have a statistical interpretation from the perspective of the field theory.  For example, on a flat spacetime, $-\mathrm{tr}(\rho \ln \rho)$ should be the same regardless of the value of $\xi$.  On a curved spacetime, the statistical entropy may be affected, but one would expect this dependence to have the same form as from the corresponding position dependent mass term $m^2 \phi^2$ where $m^2 = R$.\footnote{Of course, if the QFT is merely an effective field theory coming from a UV complete quantum gravity theory, then such terms will presumably have a statistical interpretation in terms of the underlying quantum gravity microstate counting, just like the Bekenstein-Hawking entropy term which is proportional to the area.  Our point is that there is no statistical interpretation in the QFT regime.}

%For simplicity, we also assume that appropriate field-redefinitions have been made to remove \AW{Sentence needs to be completed...}

We start by considering all renormalizable scalar field theories in the range $2 \ge d \ge 6$ and analyzing the state-dependent divergences that may appear in the entanglement entropy.  Our results are summarized in the following chart:

%STATE-DEPENDENT DIVERGENCES IN FLAT SPACETIME
%
%DIMENSION     MARGINAL POWER    BEST ODD	STATE DEP    	      DIVERGENCE
% 2		phi^n (2)	   "		phi^n (0)		 log		
% 3		phi^6 (3)	phi^5 (2.5)	phi (.5), phi^2 (1)	 log
% 4		phi^4 (4)	phi^3 (3)	phi (1), phi^2 (2)	 log	
% 5		   -		phi^3 (4.5)	phi (1.5)		linear
% 6		phi^3 (6)	   "		phi (2)			 quad

%state-dependent Divergences in Renormalizable Theories
%
%DIM      MARGINAL         BEST ODD	STATE DEP    	      DIVERGENCE
%2	\varphi^n (2)	   "		\varphi^n (0)		 log		
%3	\varphi^6 (3)	\varphi^5 (2.5)	\varphi (.5), \varphi^2 (1)	 log
%4	\varphi^4 (4)	\varphi^3 (3)	\varphi (1), \varphi^2 (2)	 log	
%5	   -		\varphi^3 (4.5)	\varphi (1.5)		linear
%6	\varphi^3 (6)	   "		\varphi (2)		 quad

\begin{center}
Table 1: State-Dependent Divergences in Renormalizable Scalar Theories
\begin{centering}
\begin{tabular*}{.75\textwidth}{@{\extracolsep{\fill}}|c|c|c|l|c|}
  \hline
\footnotesize{DIM} & \footnotesize{MARGINAL} & \footnotesize{BEST ODD} &
\footnotesize{STATE DEPEND} & \footnotesize{DIVERGE} \\
  \hline
2& $\! f(\varphi)(\nabla \varphi)^2 \!$ & $\! f_\mathrm{odd}(\varphi)(\nabla \varphi)^2 \!$ & $\langle g(\varphi) \rangle$ (0) & $\log \Lambda$ \\
  \hline
3& $\varphi^6$ & $\lambda \varphi^5$ (\textonehalf) &
$\lambda \langle \varphi \rangle$, $\langle \varphi^2 \rangle$ (1) & $\log \Lambda$ \\
  \hline
4& $\varphi^4$ & $\lambda \varphi^3$ \!(1) &
$\lambda \langle \varphi \rangle$, $\langle \varphi^2 \rangle$ (2) & $\log \Lambda$\\
  \hline
5& -- & $\lambda \varphi^3$ (\textonehalf) & $\lambda \langle \varphi \rangle$ (2), $\lambda^3 \langle \varphi \rangle$ (3) & $\!\Lambda, \log \Lambda\!$ \\
  \hline
6& $\varphi^3$ & $\varphi^3$ &
$\phantom{\lambda} \langle \varphi \rangle $ (2), $\langle X \varphi \rangle$ (4) & $\!\Lambda^2, \log \Lambda\!$ \\
  \hline
\end{tabular*}
\end{centering}
\vspace{-5pt}
\begin{eqnarray}
 X = \varphi, \,\, m^2, \,\,
 {\cal R},\,\, g^{ab}_\perp\,{\cal R}_{ab} + (K^i_i)^2/2,\,\,\textbf{or}\,\,{\cal R}_{abcd} \epsilon^{ab} \epsilon^{cd} + K_{ij} K^{ij}.
\end{eqnarray}
\end{center}
%CHECK SIGNS

\noindent The first column DIM indicates the spacetime dimension.  (As is well known, there are no perturbatively renormalizable interactions in $d > 6$.)

The MARGINAL column indicates the number of powers of $\varphi$ in the action which are necessary to make a marginally renormalizable interaction term, with the weight of the term being of course equal to the dimension.  In order to count as an interaction term, it must depend on more than two powers of the field $\phi$.  For $d=2$, the marginal interactions depend on derivatives of the scalar field, and take the form of a nonlinear sigma model interaction.  In $d = 5$ no marginal term is possible.  These marginal interactions can be used to construct the most divergent possible Feynman diagrams that appear in the effective action.\footnote{We have not assigned a variable name to the coupling that sources marginal interactions, but it should be understood that the state dependent divergences will be complicated functions of this coupling, since a Feynman diagram can be decorated by an arbitrarily large number of marginal vertices without changing its degree of divergence.}

The BEST ODD column indicates the highest weight renormalizable term with an \emph{odd} number of scalar fields in it.  This is important because no Feynman diagram can have an odd number of external legs, unless it contains at least one vertex with an odd number of legs.  Thus, in order to obtain a $\langle \varphi \rangle$-dependent divergence in $d = 3,4,5$, we must include a relevant vertex in the Feynman diagram.  We have written the relevant source term which turns on the interaction as $\lambda$, and put the dimension of $\lambda$ in parentheses next to the interaction term.

In $d = 2,6$ there are already terms with an odd number of vertices among the marginal couplings; in this case we have copied the term from the previous column into this one.  In the case $d = 2$, we require $f$ to have a piece which is odd under $\varphi \to -\varphi$ in order to get an odd state dependent divergence.

STATE DEPEND indicates which expectation values of the scalar field may appear as coefficients in an entropy divergence.  (In the case $d = 2$, the state-dependent divergence is written as $g(\phi)$ to emphasize that this is not the same function $f(\phi)$ which appears in the action.)  The quantity in parentheses represents the weight of the scalar object listed in the entropy, including any relevant source terms.  In order for there to be a divergence, one needs to find a divergent term in the effective action proportional to the quantity indicated times $R$, the Ricci scalar.  Upon performing the replica trick, the $R$ drops out, subtracting 2 from the weight.  Hence in order to be a divergence, the weight of the terms listed must be no more than $d - 2$.

Finally, DIVERGE indicates the maximum degree of divergence of the expressions in the previous column.  This is calculated by subtracting the weight of the state-dependent divergence from $d-2$.
%In the case of a $\langle \phi^2 \rangle$-divergence, this is calculated by subtracting the weight of $\phi^2 R$ from the dimension $d$.  In the case of a $\langle \phi \rangle$-divergence, one subtracts the weight of $\phi R$ from the weight of the density in the BEST ODD column.

In the case of $d = 6$, in addition to the quadratic $\langle \varphi \rangle$-divergence, it is also possible to have a log divergence, by replacing the $X$ in the expression $\langle X \varphi \rangle$ with any of five possible weight 2 items, listed below the chart.  Here $g^{ab}_\perp$ is the inverse metric normal to the entangling surface, and $i, j$ are indices restricted to the entangling surface.  These terms come from the renormalization of the following terms in the action respectively:
\begin{equation}
\varphi^2 R,\,\,
m^2 \varphi,\,\,
\varphi R^2,\,\,
\varphi (R_{ab})^2,\,\,
\varphi (R_{abcd})^2,
\end{equation}
and we have used the work of \cite{FPS,Dong:2013qoa,Camps:2013zua} to determine the extrinsic curvature dependence of the entropy for the final two cases.

A sixth candidate term $\nabla^2 \varphi R$, which produces a $\nabla^2 \varphi$ divergence, is not listed either here or above, since it can be related to the other terms by means of the field equation for $\phi$ (or equivalently, by a field redefinition of the action).  A seventh candidate term $\nabla_a \nabla_b \varphi G^{ab}$ is a total derivative as a result of the Bianchi identity, and hence should not contribute.\footnote{Its Wald entropy \cite{Wald:1993nt}, obtained by differentiating with respect to the Riemann curvature, is proportional to $g^ij \nabla_i \nabla_j \phi$, with $i,j$ restricted to the four dimensional entangling surface.  However due to the ambiguities in the Noether charge approach, this formula is only valid for stationary entangling surfaces \cite{Jacobson:1993vj,Iyer:1995kg}.  Presumably once the extrinsic curvature terms are properly calculated a la \cite{FPS,Dong:2013qoa,Camps:2013zua}, one finds that it is actually $g^ij D_i D_j \varphi G^{ab}$ where $D_a$ is the covariant derivative intrinsic to the geometry of the entangling surface.  This is a total derivative, and thus does not contribute to the entropy of a compact entangling surface.}

For $d \le 4$, there also exist peturbatively renormalizable interactions involving spinor or vector fields.  These add additional possible marginal and relevant interactions.  However, it turns out that they do not add new kinds of state-dependent divergences, for the following reasons:

In $d = 2$ any state-dependence in the entropy must be weight 0.  The only possible terms with weight 0 are functions of scalar field $\phi$, which are already included.  Such terms are already allowed in pure scalar field theories.  Allowing additional marginal terms in the action may change the coefficients of the state dependent divergences, but not the allowed kinds of divergences.

Since a spinor bilinear $\langle \psi_1 \psi_2 \rangle$ is weight $d - 1$, it cannot appear as a coefficient of a divergence.  So we can only consider diagrams where spinors are internal lines.  Such interactions do not introduce any qualitatively new kinds of state dependent divergences.  Spinor interactions may add new kinds of relevant source terms, but they do not change the set of field operators which can appear in state-dependent divergences.

One might have thought that in $d = 3, 4$, the Yukawa couplings might help by producing diagrams with an odd number of external $\varphi$ lines.  But in $d = 3$, the Yukawa coupling has dimension 2\textonehalf, so it is no better than $\varphi^5$, while a marginal $\varphi^2 \psi^2$ coupling produces an even number of scalars.  While in $d = 4$, the Yukawa couplings (and all other marginal couplings) preserve an accidental $\mathbb{Z}_2$ symmetry that counts the number of scalars plus left-handed fermions, mod 2; this prevents any new kinds of state-dependent divergences from existing.\footnote{This assumes that the matter number is conserved so that left-handed matter can be consistently distinguished from left-handed antimatter; otherwise one must instead use the accidental $\mathbb{Z}_4$ symmetry given by $N_L - N_R + 2N_\mathrm{scalars}$ mod 4.  Note also that since our analysis is perturbative we are neglecting anomalous instanton effects.}

A minimally coupled gauge boson also does not change anything of significance.  In $d = 4$, on dimensional grounds one might have expected divergences proportional to either the electric flux $\langle F_{ab} \rangle \epsilon^{ab}$ or the magnetic flux $\langle *F_{ab} \epsilon^{ab} \rangle$, in C-violating theories such as the Standard Model.  But in fact these terms are ruled out by covariance, since there is no way to contract one copy of $F_{ab}$ with the Riemann tensor to build a viable term in the action.  Furthermore, $F_{ab}$ has dimension $d/2$, and thus cannot appear directly in perturbatively renormalizable interactions with scalars or spinors.

A massive vector boson $V_a$ is also not useful, since the longitudinal modes of $V_a$ also have weight $d/2$ for purposes of renormalization theory.  In $d = 3,4$, the only new perturbatively renormalizable terms are the Proca mass $V^a V_a$ (which is not useful), and the mixed propagator $V^a \nabla_a \varphi$ (which can be removed by a field redefinition).

Interactions involving higher spin fields are necessarily nonrenormalizable in $d > 2$, so we conclude that the table above gives a complete list of the possible state-dependent divergences.

%%%%%%%%%%%%%%%%%%%%%%%%%%%%%%%%%%%%%%%%
%
\section{Holographic Theories}
%
%%%%%%%%%%%%%%%%%%%%%%%%%%%%%%%%%%%%%%%%
\label{strong}

Having classified all state-dependent entropy divergences that can arise in perturbation theory around a free fixed point, it is natural to ask about more general theories where the interactions may be strong.  Consider, for example, covariant relevant and marginal deformations of unitary conformal field theories with asymptotically AdS gravity duals.
We will work at the level of classical bulk physics, or equivalently at leading order in a limit where an appropriate integer $N$ labelling the holographic field theory has been taken to be large.

As discussed in the introduction, entropy divergences are constrained only by their connection to action divergences via \eqref{replica}.  In the holographic context, this point was recently emphasized by \cite{Taylor:2016aoi} in connection with the Lewkowycz-Maldacena argument \cite{Lewkowycz:2013nqa} for the Ryu-Takayanagi (RT) \cite{Ryu:2006bv} and covariant Hubeny-Rangamani-Tayakanagi (HRT) \cite{Hubeny:2007xt} entropies.  This in particular means that the degree of divergence will be $d-\Delta$, where $\Delta \le d$ is the weight of the corresponding term in the action.  From \eqref{replica}, action terms that are algebraic in the metric give no contribution to the entropy.  As a result, we will show below that non-trivial contributions arise only from terms containing two or more derivatives of the metric. For example, the presence of a scalar operator ${\cal O}$ of dimension $\Delta_{\cal O} \le d-2 $ generally requires an action counter-term of the form
\begin{equation}
\label{OR1}
\int d^d X R {\cal O}
 \end{equation}
 and so (as in section \ref{pertclass} for ${\cal O} = \varphi, \varphi^2$) indicates an entropy divergence of degree $d-2 - \Delta_{\cal O}$ proportional to the integral of ${\cal O}$. We consider only cases with $d \le 6$.  Recall that there are no known holographic theories no known examples of interacting conformal field theories (holographic or otherwise) with $d \ge 7$. With this restriction, we will see that all possible action counter-terms take the form \eqref{OR1}.

For our purposes, the key feature of large $N$ holographic theories that they  admit a description via a weakly-coupled bulk path integral, inside which the dimensions of composite operators are given by simply adding the dimensions of their component `elementary' operators and sources.  In this context, let us use the term `operator' below to refer only to objects for which the dependence on the background metric is at most algebraic when the elementary operators  are held fixed.  With this understanding, boundary terms terms in the bulk gravitational action contribute to \eqref{replica} only when they involve {\it explicit} dependence on derivatives of the metric.  Since we in any case integrate over all values of the elementary operators, there is no effect from any implicit dependence of these operators on the background metric.\footnote{In the bulk semiclassical approximation, the corresponding statement is that the requirement for the full bulk action be stationary requires boundary term contributions from changes in operators ${\cal O}$ with $n$ to cancel against other contributions from the bulk.}

The only other property of holographic theories used below is that while unitarity generally requires the dimension of scalar operators ${\cal O}$ to satisfy only $\Delta_{\cal O} \ge (d-2)/2$, in ghost-free holographic theories they in fact satisfy the strict inequality $\Delta_{\cal O} > (d-2)/2$ \cite{Andrade:2011dg}.  This is to be expected as unitary generally allows $\Delta_{\cal O} =(d-2)/2$ only when the correlators of ${\cal O}$ are those of a free field.

Now, since we require at least one derivative of the metric, the operator in our action term can have dimension at most $d-1$.  But the unitarity bounds (see e.g. \cite{Minwalla:1997ka} and references therein) require any operator with dimension $d-1$ or less to be a scalar,  a derivative of a scalar, an antisymmetric tensor, or a conserved vector (satisfying $\nabla_a j^a =0$).   We may ignore the anti-symmetric tensors as they cannot be combined with derivatives or powers of the Riemann tensor to build a covariant term.  And since we forbid vector sources, any conserved vector operator can appear only through $\nabla_a j^a =0$.

This leaves us with terms that involve only scalar operators ${\cal O}$.  Integrating by parts allows us to remove derivatives from ${\cal O}$, so it suffices to consider only terms given by multiplying such a scalar ${\cal O}$ by a scalar $\Phi$ built from the metric.  The possible such terms are then classified by scalars $\Phi$ of weight $d - \Delta_{\cal O}$ or less\footnote{The same argument applies when one allows position-dependent scalar sources of non-negative conformal weight, though then $\Phi$ can depend on these scalar sources as well.}.

Since $\Delta_{\cal O} > (d-2)/2$, the scalar $\Phi$ can contain at most three derivatives.  This in particular forbids divergences associated with terms that the table in section \ref{pertclass} would describe as having $\ln \Lambda$ divergences for $d=6$.  Covariance requires derivatives to occur in pairs, so only one pair can be present.  In this case the only allowed action-counterterm than can affect the entropy is just \eqref{OR1} as claimed above.

%%%%%%%%%%%%%%%%%%%%%%%%%%%%%%%%%%%%%%%%
%
\subsection{State-dependent RT divergences}
%
%%%%%%%%%%%%%%%%%%%%%%%%%%%%%%%%%%%%%%%%
\label{holo}

We suggested above that action divergences of the form \eqref{OR1} should be generic in leading-order large $N$ holographic theories when $\Delta_{\cal O} \le d-2$ in the presence of low dimension sources, and that they should be accompanied by state-dependent divergences in the RT and HRT entropies.   Although the literature contains statements \cite{Hung:2011ta} that such divergences do not in fact occur, we now describe bottom-up examples where they do, and which support our claim that they are generic.  We will analyze the entropy divergences directly, though a similar computation may of course be performed at the level of the action. As above, we will work at the level of classical bulk physics, or equivalently at leading order in a limit where an appropriate integer $N$ labelling the holographic field theory has been taken to be large.

It is sufficient to take the bulk dual to consist of $(d+1)$-dimensional Einstein-Hilbert gravity with negative cosmological constant coupled to two scalar fields $\phi, \chi$.  We will study solutions locally asymptotic to AdS$_{d+1}$, suppressing discussion of any possible compact factor $X$ in the bulk space time (though in a top-down model the scalars $\phi,\chi$ may in fact arise from Kaluza-Klein reduction on $X$).  We take the bulk scalar action to be of the standard second-derivative form with scalar potential

\begin{equation}
\label{V}
V(\phi, \chi) = \frac{1}{2} m^2_\phi \phi^2 + \frac{1}{2} m^2_\chi \chi^2 + g(\chi) \phi + \dots,
\end{equation}
where $\dots$ indicates terms of at least third order in $\phi$. We assume that $g(\chi)$ admits a power series expansion about $\chi=0$, and whose first non-trivial terms are
\begin{equation}
g(\chi) = \alpha \chi^n + \tilde \alpha \chi^{\tilde n} + \dots
 \end{equation}
 with $\tilde n > n \ge 2$.  Taking the AdS length scale to be $\ell=1$, the bulk field $\chi$ is associated with a source $s$ of conformal weight $\Delta_s$ and an associated operator of weight $d - \Delta_s$ determined by $m^2_\chi =-4 \Delta_s(d-\Delta_s)$, perhaps with additional input from a choice of boundary conditions.  The bulk field $\phi$ is similarly associated with an operator ${\cal O}$ of conformal weight $\Delta_{\cal O}$ as well as an associated source of weight $d - \Delta_{\cal O}$, with $\Delta_{\cal O}$ determined by $m^2_\phi =-4 \Delta_{\cal O}(d-\Delta_{\cal O})$ and perhaps again a choice of boundary conditions.  To avoid ghosts \cite{Andrade:2011dg}, we choose $\Delta_{\cal O} > (d-2)/2$.  To preserve the asymptotically AdS boundary conditions we require $\Delta_s \ge 0$. We will not need to turn on the source for ${\cal O}$, and it turns out that we will be interested only in $\Delta_s < \frac{d+2}{4} \le d/2$ where the last inequality uses $d \ge 2$.  For simplicity, we also require $\tilde n \Delta_s > \Delta_{\cal O}$. Our bulk scalars then admit asymptotic expansions
\begin{equation}
\label{aspc}
\phi = \sum_{ k,m \ge 0} \alpha^{1+2m} z^{((n+2(n-1)m)\Delta_{s} + 2k) }P_{k,n+2m}(\nabla,s) - \frac{ z^{\Delta_{\cal O}}}{d-2\Delta_{\cal O}} {\cal O} + \dots,
\end{equation}
\begin{equation}
\label{chiexp}
\chi = \sum_{ k,m \ge 0} \alpha^{2m} z^{((1+2(n-1)m)\Delta_{s} + 2k) }Q_{k,1+2m}(\nabla,s)  + z^{((n-1)\Delta_{s} + \Delta_{\cal O}) } \alpha \gamma  s^{n-1} {\cal O}  + \dots,
\end{equation}
in terms of a Fefferman-Graham radial coordinate $z$ (see below).
In the above, $+\dots$ represents terms of higher order in $z$ than the last term explicitly displayed and
$P_{x,y},Q_{x,y}$ are scalar polynomials in $s$ and its derivatives in the QFT spacetime (i.e., derivatives along the boundary directions from the bulk point of view) which are homogeneous of order $x$ in derivatives and order $y$ in $s$.  When $(n+2(n-1)m)\Delta_{s} + 2k = \Delta _{\cal O}$,
the $P_{n+2m, k}(\nabla,s)$ term will also contain a factor of $\ln z$, but any other logs must be a part of the higher order terms indicated by $+\dots$.   See e.g. \cite{Amsel:2006uf,Amsel:2007im} for the details of an analogous computation and e.g. \cite{Marolf:2006nd} (as well as the earlier work \cite{Minces:2001zy,Mueck:2002gm,Minces:2002wp,Sever:2002fk}) for a discussion of the normalization of the coefficient of ${\cal O}$ in \eqref{chiexp}.  For our purposes, the important coefficients and polynomials are
\begin{equation}
\label{QPs}
Q_{0,1} = 1,  \   \ P_{0,n} = \alpha \sigma s^n, \ {\rm for} \  \sigma =  \frac{1 }{n \Delta_s (\Delta_s - d) + \Delta_{\cal O}(d-\Delta_{\cal O}) },
\end{equation}
\begin{equation}
\label{gamma}
{\rm and} \ \gamma = -\frac{ n }{(d-2\Delta_{\cal O}) \left[(\Delta_{\cal O} + (n-1)\Delta_s)(\Delta_{\cal O} + (n-1) \Delta_s) -d  + \Delta_s(d-\Delta_s) \right] }.
\end{equation}

The RT and HRT conjectures state that the (divergent) entropy of the field theory restricted to a region of its spacetime is given by the area of an appropriate bulk extremal $(d-1)$ surface anchored to the asymptotically locally AdS boundary on some $(d-2)$ surface.  Such divergences are dictated by the asymptotic expansion of the bulk metric ${\sf g}_{AB}$, which is usefully expressed in the Fefferman-Graham gauge in terms of bulk coordinates $X^A = (z, x^a)$ as

\begin{equation}
\label{GFG}
{\sf g}_{AB} dX^A dX^B = \frac{1}{z^2} \[ dz^2 + g_{ab}(z)  dx^a dx^b \],
\end{equation}
with  $\lim_{z \rightarrow 0} g_{ab}(z)$ giving the metric $g^{(0)}_{ab}$ of the spacetime with coordinates $x^a$ carrying the $d$-dimensional holographic QFT. Indeed, \cite{Graham:1999pm} shows that HRT divergences are determined by the terms in \eqref{GFG} of order $z^{d-2}$ or larger as $z \rightarrow 0$.\footnote{Ref. \cite{Graham:1999pm} analyzed only the so-called universal sector of holographic CFTs, in which the Fefferman-Graham expansion of the metric at these orders does not involve bulk matter fields.  But it is clear from \cite{Graham:1999pm} that the conclusion holds more generally.}  This is particularly clear in contexts with sufficient symmetry to guarantee the extremal surface to be described by fixing the values of two of the $x^a$ coordinates.  The divergences are then given immediately by integrating the associated induced metric over $z$ and the remaining $x^a$, so that terms of order $z^{(d-2)}$ in \eqref{GFG} induce logarithmic divergences.  By our general arguments above, at least for $2 \le d \le 6$ such highly symmetric cases are sufficient to determine the coefficient of all allowed divergences.

The expansion of $g_{ab}(z)$  may be computed by iteratively solving the $(A,B) = (a,b)$ components of the bulk Einstein equation,
\begin{equation}
\label{EE}
G_{AB}  -\frac{d(d-1)}{2} {\sf g}_{AB} = 8 \pi T^{bulk \ matter}_{AB},
 \end{equation}
 where $G_{AB}$ is the bulk Einstein tensor and, since $\ell=1$ the factor $\Lambda = -\frac{d(d-1)}{2}$ is the bulk cosmological constant.  On the right-hand side,
\begin{equation}
\label{bulkmatter}
T^{bulk \ matter}_{AB} = D_A \phi D_B \phi +  D_A \chi D_B \chi - \frac{1}{2} {\sf g}_{AB} \left(   D_C \phi D^C \phi +  D_C \chi D^C \chi +2 V \right).
\end{equation}
is the bulk matter stress tensor in terms of the bulk covariant derivative $D_A$.  As described in \cite{Hollands:2005ya}, when one disentangles the iterative equation, it turns out to be the trace-reversed bulk stress tensor
\begin{equation}
\label{L}
L_{AB} = T_{AB} - \frac{{\sf g}_{AB}}{d-1} {\sf g}^{CD}T_{CD}
\end{equation}
whose $ab$ components feed directly into the recursion relation for coefficients in the expansion of $g_{ab}(z)$.

For $T^{bulk \ matter}_{AB} = 0$ the resulting expansion takes a familiar form that at order $z^{d-2}$ or less involves only even powers of $z$ and no logarithms, with coefficients dictated by the field theory's metric $g^{(0)}_{ab}$.  But in the presence of scalar sources and operators of sufficiently low dimension, the $T^{bulk \ matter}_{AB}$ on the right hand side of \eqref{EE} can also contribute.  Indeed, from \eqref{aspc}, \eqref{bulkmatter}, and \eqref{L}, we see that for small enough $\Delta_s$ the lowest order term involving ${\cal O}$ in $L_{ab}$ arises from a combination of i) the $g(\chi)\phi$ term in \eqref{V}, ii) the $m^2_\phi$ and $m^2_\chi$ terms in $V$, and iii)  the $\partial_z$-parts of the kinetic terms in $T^{bulk \ matter}_{AB}$.  All contributions are of the form $z^{n\Delta_s + \Delta_{\cal O}} g^{(0)}_{ab} s^n {\cal O}$, giving a term $\alpha\beta z^{n\Delta_s + \Delta_{\cal O}} g^{(0)}_{ab} s^n {\cal O}$ in $g_{ab}(z)$.  A straightforward calculation gives
\begin{equation}
\label{beta}
\beta = \frac{2[(d-1)K + dW]}{(d-1)(n\Delta_s + \Delta_{\cal O})(n\Delta_s + \Delta_{\cal O} -1)},
\end{equation}
with
$K$ and $W$ representing respectively the contributions from the terms involving $\partial_z$ and those algebraic in bulk fields. We find
\begin{eqnarray}
K &=& \frac{1}{2}\left( -\frac{\sigma n \Delta_{\cal O} \Delta_s}{(d-2\Delta_{\cal O})}  +\gamma \Delta_s \left[\Delta_{\cal O} + (n-1) \Delta_s \right] \right) \cr
&=& - \frac{n (n-1) \Delta_s^2}{\Delta_{\cal O} (d-\Delta_{\cal O}) (d-2\Delta_{\cal O})} \left(1 + O(\Delta_s) \right)
\end{eqnarray}
and
\begin{eqnarray}
\label{WK}
W &=& \frac{\sigma \Delta_{\cal O} (d-\Delta_{\cal O}) - 1}{(d-2\Delta_{\cal O})} - \gamma \Delta_s(d-\Delta_s) .
\cr
&=&\frac{n \Delta_s^2 [d^2 + \Delta_{\cal O} (d-\Delta_{\cal O}) + d(n-1) (2 \Delta_{\cal O}-d) + nd^2] } {(d-2\Delta_{\cal O}) \Delta_{\cal O}^2 (d-\Delta_{\cal O})}
\left( 1 + O(\Delta_s) \right),
\end{eqnarray}
where $\gamma, \sigma$ were given in \eqref{QPs}, \eqref{gamma}.

For $\beta \neq 0$ this term yields a state-dependent divergence of order $(d-2) - n\Delta_s + \Delta_{\cal O}$.  Here order zero represents a logarithm and a negative result indicates no divergence.  Since $\Delta_{\cal O}$ can be close to $(d-2)/2$ and $\Delta_s$ can be close to zero, our class of models leaves ample room for non-negative degrees of divergence. It is clear that for $\Delta_s \neq 0$ there is generically no cancellation between the $K$ and $W$ terms in \eqref{beta}, and that the leading divergence is unchanged by adding additional terms to $V$, including $\phi$-independent terms proportional to $\chi^{\hat n}$ for $\hat n > 2$.

In contrast, our results \eqref{WK} vanish quadratically near $\Delta_s =0$.
But since it becomes increasingly difficult to ignore additional terms in this regime, it remains an open question whether state-dependent divergences are allowed for vanishing  $\Delta_s$.  It would thus be interesting to explore the $\Delta_s=0$ case further.  This is particularly so as the proto-typical holographic theory of $4d$ ${\cal N}=4$ SU(N) super-Yang-Mills (dual to type IIB supergravity compactified to AdS$_5 \times S^5$) has the property \cite{Kim:1985ez} the lowest dimension operator has $\Delta_{\cal O}=2$ so that any sources in a term of dimension $d-2=2$ or less must have $\Delta_s=0$.  An analogous statement holds for eleven-dimensional supergravity compactified to AdS$_4 \times S^7$ (where $d=3$ and the lightest scalar operator has $\Delta_{\cal O}=1$), though we have not surveyed more general top-down models of holographic theories.

\section{Finiteness of Various Quantities}\label{finite}

Although the entanglement entropy may have state-dependent divergences, there are several closely related quantities in which all divergences are expected to cancel (including therefore state-dependent divergences).  These include three closely related quantities: the mutual information, the generalized entropy, and the relative entropy:

\paragraph{Mutual Information} State-dependent divergences cannot afflict computations of mutual information $I(A:B) =  S(A) +S(B) - S(A\cup B)$, when the regions $A$ and $B$ are separated by a finite proper spatial distance.  This is because all divergences cancel between the various boundary regions \cite{Casini:2006ws}.  

Note that the mutual information is a special case of the relative entropy: $I(A:B) = S(\rho_{AB}\,|\,\rho_A \otimes \rho_B)$.

\paragraph{Generalized Entropy} On a similar note, even in the presence of our state-dependent divergences, coupling our quantum field theory to gravity yields a finite generalized entropy $S_{gen} = S_\mathrm{BH} + S_{outside}$, where $S_{BH}$ is the entropy of the black hole including any state-dependent counterterms.  Due to the replica trick, this follows directly from finiteness of the renormalized partition function.  (The renormalization procedure has been extensively studied in the literature; see the Appendix of \cite{Bousso:2015mna} for a review and citations.)

In cases where a QFT state makes a small gravitational perturbation to a Killing horizon, the generalized entropy on the causal horizon is given by $S_\text{gen} = C - S(\rho\,|\,\sigma)$ where $\sigma$ is the associated Hartle-Hawking state and $C$ is an additive constant \cite{Casini:2008cr,Wall:2010cj,Wall:2011hj}.

\paragraph{Relative Entropy} The fact that relative entropy $S(\rho\,|\,\sigma)$ is finite (for well-behaved states) should also be confirmable by replica trick calculations; we will now show this in cases where, for simplicity, the states $\rho$ and $\sigma$ both come from a path integral which is rotationally symmetric around the entangling surface $\partial R$.  To get a divergence that depends on the state, we assume that some scalar $\Phi$ (e.g. $\langle \phi \rangle$ or $\langle \phi^2 \rangle$) associated with the state-dependence differs between the two states at the entangling surface (due to some rotationally symmetric source or boundary condition), so that $S(\rho) - S(\sigma)$ is infinite.

Let us now consider the path integral formed by gluing together $r$ consecutive copies of the path integral used to define $\rho$, with $s$ consecutive copies of the path integral used to define $\sigma$, for a total angle deficit at the origin of $2\pi(1 - r - s)$.  This path integral defines the partition function $Z(r,s) = \mathrm{tr}(\rho^r \sigma^s)$, where $\rho$ and $\sigma$ are not yet taken to be normalized.  Since the whole setup is rotationally symmetric, we can allow $r$ and $s$ to take noninteger values and still retain a geometrical description.

The modular Hamiltonian $K = \ln \sigma$ for the state $\rho$ is now given by
\begin{equation}
K(\rho) = -\partial_s \ln Z(1,0),
\end{equation}
while the entropy (after normalization of $\rho$) is given by
\begin{equation}
S(\rho) = (1 - \partial_r) \ln Z(1,0).
\end{equation}
Hence the relative entropy is
\begin{equation}\label{relS}
S(\rho\,|\,\sigma) = \Delta K - \Delta S = 
(\partial_r - \partial_s - 1) \ln Z(1,0) - \ln Z(0,1),
\end{equation}
where the first two terms require differentiating with respect to a small conical angle deficit, while the last two terms are evaluated on the original smooth space time.

Let us assume that in order to properly define the modular Hamiltonian $K$ above, all bulk divergences of $\ln Z$ \emph{not} associated with the conical angle deficit have already been renormalized by absorption into bulk counterterms.  We therefore restrict attention to the divergences which multiply the conical angle deficit $1 - r - s$, appearing in the first two terms.  For example, these might correspond to state-dependent divergences which are absorbed by a nonminimal $\Phi {\cal R}_\text{sing}$ term, where ${\cal R}_\text{sing}$ is the singular part of the curvature associated with the conical singularity.\footnote{If the background manifold has ${\cal R} \ne 0$, there may of course also be a nonsingular contribution from nonminimal coupling terms, but divergent contributions to this term are already renormalized in defining $K$, as stated above.}

It seems reasonable to suppose that the scalar quantity $\Phi$ at the singularity should itself be a smooth function of $r$ and $s$ at $\ln Z(1,0)$, namely $\Phi(r,s)$.  Then at first order, the state-dependent divergence in the effective action is given by
\begin{equation}\label{Taylor}
\ln Z_\mathrm{div} \propto (1 - r - s) \Phi(1,0)
\end{equation}
where the $\partial_r$ and $\partial_s$ derivatives in the Taylor expansion cannot act on $\Phi$ because  ${\cal R}_\text{sing} = 0$ at $\ln Z(1,0)$.  But combining the derivative terms from \eqref{relS} with \eqref{Taylor},
\begin{equation}
(\partial_r - \partial_s) (1 - r - s) = 0;
\end{equation}
Therefore there are no state-dependent divergences in the relative entropy between the two states $\rho$ and $\sigma$, defined by the path integrals above.  The underlying reason is that the relative entropy \eqref{relS} can be evaluated using only the nonsingular partition function at $r + s = 1$.  However, state-dependent divergences may still be present if we consider $\Delta S$ or $\Delta K$ on their own.

More generally, we may consider states $\rho$ and $\sigma$ defined by non-rotationally symmetric path integrals.  Here we lose the geometrical interpretation, but we expect that for purposes of analyzing the UV divergence structure near the entangling surface, a formally similar argument will still hold.

\section{Discussion}
\label{disc}

By working in both perturbative and holographic contexts, we have shown that the von Neumann entropy of a field theory in a region of spacetime can display a variety of state-dependent divergences.  Each such divergence is associated with divergences in the (bare) partition function involving the Ricci or Riemann curvature of the background spacetime, with the possible such terms classified by the low-dimension scalar operators present in the theory. We have argued that the coefficient of such divergences is generically non-zero, but it remains possible that state-dependent divergences are forbidden in theories with exact conformal symmetry.  We also remind the reader that our holographic examples were constructed by simply postulating a certain bulk Lagrangian.  It thus remains to be shown that the required structure actually arises in models with known field theory duals.

The state-dependent logarithmic divergences are particularly interesting because, as usual, any compensating logarithmic counter-term in the action requires a choice of scale.  As a result, such terms constitute a new type of conformal anomaly and provide a corresponding state-dependent contribution to the trace of the stress tensor.

Furthermore, a priori, there is no preferred choice of the scale in this counter-term.   In curved space, we note that the choice of scale directly affects the correlation functions of the theory.  But since the term is proportional to some power of Ricci or Riemann curvature, this is not so in flat spacetime.  In that case, taking the correlators to define the QFT means that the choice of scale can have no physical effect.  Nevertheless, changing the scale will shift the renormalized entropy $S_{ren}$ by some finite amount.

This argues that quantum field theories generally have only families of finite quantities that could be called renormalized von Neumann entropy, but that there is no preferred member of this class.  The same issue has been raised and discussed many times before in the context of possible finite (i.e., non-divergent) curvature couplings (see e.g. \cite{Larsen:1995ax,Solodukhin:1996jt,Hotta:1996cq,Solodukhin:2011gn,Klebanov:2012va,Nishioka:2014kpa,Lee:2014zaa,Herzog:2014fra,Rosenhaus:2014ula,Dowker:2014zwa,Casini:2014yca,Rosenhaus:2014zza}). In that context one might hope (as in some of the above references) that either minimal coupling or some other prescription will give rise to a preferred definition of von Neumann entropy, but in cases with a logarithmic divergence any such prescription must entail the introduction of a new preferred scale.

On the other hand, we have argued that certain universal quantities like relative entropy $S(\rho\,|\,\sigma)$ should remain finite and independent of the above ambiguity.  When $\sigma$ is the Hartle-Hawking state associated with a bifurcate Killing horizon, this could also be derived by noting that the computations of boundary terms are essentially classical and then using the argument of \cite{Fursaev:1998hr} to show cancellation between ``energy'' and ``entropy'' contributions.

As a final comment, we note on the other hand that more general state-dependent entropy divergences will generally arise in effective low-energy field theories as these generally feature couplings or background fields with negative mass dimensions.  In such contexts, the action counter-terms may contain arbitrary operators multiplied by any number of derivatives and powers of the curvature, leading to correspondingly complicated state-dependent divergences in the entropy.  However, if the theory flows to a UV fixed point, these extra divergences will be regulated by the short-distance physics.

\section*{Acknowledgements}
We thank David Berenstein, Nathaniel Craig, William Donnelly, Tom Hartman, Rob Myers, Joe Polchinski, Vladimir Rosenhaus,  and Mark Srednicki for useful discussions.  DM was supported in part by the Simons Foundation and by funds from the University of California.  AW is supported by the Institute for Advanced Study, the Martin A. and Helen Chooljian Membership Fund, and NSF grant PHY-1314311.

\bibliographystyle{JHEP}
\bibliography{SD}

\end{document}